\documentstyle[preprint,12pt,aps]{revtex}

\begin{document}
\draft
\title{Cluster Heat Bath Algorithm in Monte Carlo Simulations of Ising Models}

\author{F. Matsubara, A. Sato, and O. Koseki }
\address{Department of Applied Physics, Tohoku University, Sendai 980-77,
Japan}
\author{T. Shirakura}
\address{Faculty of Humanities and Social Sciences, Iwate University, 
Morioka 020, Japan}

\date{13 January 1997}
%
\maketitle
\begin{abstract}
We have proposed a cluster heat bath method in Monte Carlo simulations 
of Ising models in which one of the possible spin configurations 
of a cluster is selected in accordance with its Boltzmann weight. 
We have argued that the method improves slow relaxation in complex systems 
and demonstrated it in an axial next-nearest-neighbor Ising(ANNNI) model 
in two-dimensions.
\end{abstract}



Recently various algorithms have been proposed to reduce the CPU time of 
computer in the Monte Carlo(MC) simulation\cite{Binder}. 
Cluster-flip algorithms\cite{Swendsen2,Wolff} were proposed using 
ideas from percolation theory\cite{Fortuin}. 
Although the methods were very efficient to simulate large systems near 
criticality, these were not successfully applied to complex systems 
which contain frustrated interactions such as spin-glasses. 
On the other hand, to study the ordered state of complex systems, extended 
ensemble methods were developed\cite{Swendsen1,Berg,Marinari,Hukushima}. 
In these methods, however, the conventional single-spin-flip algorithm is 
used to guarantee the ergodicity. 
Quite recently, a new update method was proposed in which the spin 
configuration of a chain of Ising spins is updated in accordance with 
the Boltzmann weight\cite{Koseki}. 
The method is very effective for quasi-one-dimensional models and enables 
us to make realistic simulations\cite{Koseki2} of quasi-one-dimensional 
Ising magnets such as CsCoBr$_3$\cite{Yelon} and CsCoCl$_3$\cite{Mekata}. 
However, it was not so useful for ordinary two- and three-dimensional 
models, especially for complex systems.

In this Letter, we propose a general configuration-update algorithm which is 
applicable to various Ising models with short-range interactions and  
very effective for improving slow relaxation in complex systems. 
We consider to find the probable spin configuration of a cluster of $N$ spins. 
This can be readily done when $N$ is small, but becomes difficult 
when $N$ is increased. 
However, if the cluster is decomposed into layers of spins and interlayer 
couplings exist only between the spins on adjacent layers, 
we can update the spin configuration of the layers step by step 
with the aid of a transfer matrix technique. Thus, we can treat a larger 
cluster, e.g., a cluster of $N \sim M \times L$ spins in a cubic lattice 
with the nearest neighbor couplings, where $L$ is the linear size of the 
lattice and $M$ is the number of the spins of the layer. 
We can prove that the spin configuration realized in this method is in 
accordance with the Boltzmann weight. 
So we call the algorithm a cluster heat bath (CHB) method. 
We argue that the CHB method improves slow relaxation in complex systems 
and demonstrate it in an axial next-nearest-neighbor Ising (ANNNI) model
\cite{ANNNI} in two-dimensions.


We start with an Ising model described by the Hamiltonian
\begin{eqnarray}
    {\cal H} = - \sum_{<i,j>}J_{ij}\sigma_{i}\sigma_{j}, 
\end{eqnarray}
where $\sigma_{i} (= \pm 1)$ are Ising spins and $J_{ij}$ are 
coupling constants. 
Now we pick out a cluster of spins and consider its probable spin 
configuration under the condition that the other spins are fixed. 
It is noted that the cluster defined here is {\it an ensemble of 
connected spins in a fixed part of the system,} not one of connected spins 
with the same sign. 
We suppose that the cluster is composed of $L$ layers and interactions exist 
only between neighboring layers. Hereafter we call the cluster as $A_L$. 
The Hamiltonian of $A_L$ is, then, described as
\begin{eqnarray}
    {\cal H}_{A_L} = - \sum_{l=1}^{L-1}H_{l,l+1} - \sum_{l=1}^LB_l ,
\end{eqnarray}
with
\begin{eqnarray}
 H_{l,l+1}&=&\sum_{i}\sum_jJ_{ij}^{(l,l+1)}
           \sigma_{i}^{(l)}\sigma_{j}^{(l+1)}, \\
 B_l &=&\sum_{ij}J_{ij}^{(l)}\sigma_i^{(l)}\sigma_j^{(l)} + 
           \sum_ih_i^{(l)}\sigma_i^{(l)},
\end{eqnarray}
where $J_{ij}^{(l)}$ and $J_{ij}^{(l,l+1)}$ are exchange interactions between 
$i$th and $j$th spins on the same layer $(l)$ and those between different 
layers $(l)$ and $(l+1)$, respectively.
The effective field $h_i^{(l)}$ of $i$th spin on the layer $(l)$ is given as 
a sum of the external field and the exchange field which comes from the spins 
surrounding $A_L$: 
\begin{eqnarray}
    h_i^{(l)} = \sum_{j}J_{ij}\sigma_{j} + mH,
\end{eqnarray}
where $m$ and $H$ are the magnetic moment and the external field, 
respectively.

Now we consider the probable spin configuration $\{\sigma_i^{(L)}\}$ of the 
layer $(L)$. 
We define the following weight function $F_l(\{\sigma_i^{(l)}\})$:
\begin{eqnarray}
 F_l(\{\sigma_i^{(l)}\}) = \sum_{\{\sigma_i^{(1)}=\pm 1 \}} \cdots 
 \sum_{\{\sigma_i^{(l-1)}=\pm 1 \}} 
 \exp(\beta\sum_{k=1}^{l-1}H_{k,k+1} + \beta \sum_{k=1}^lB_k) , 
\end{eqnarray}
where $\beta = 1/k_BT$ with $k_B$ and $T$ being the Boltzmann's constant 
and temperature, respectively. 
This function can be readily obtained from the recursion formula 
\begin{eqnarray}
 F_l(\{\sigma_i^{(l)}\}) &=& \sum_{\{\sigma_i^{(l-1)}=\pm 1 \}} 
         F_{l-1}(\{\sigma_i^{(l-1)}\})
         \exp(\beta H_{l-1,l} + \beta B_l)  \hspace{0.5cm} l \geq 2 
\end{eqnarray}
with the initial function $ F_1(\{\sigma_i^{(1)}\}) = \exp( \beta B_1)$. 
The probability $P_L(\{\sigma_i^{(L)}\})$ of the spin configuration 
$\{\sigma_i^{(L)}\}$ of the layer $(L)$ is given as 
\begin{eqnarray}
 P_L(\{\sigma_i^{(L)}\}) = \frac{F_L(\{\sigma_i^{(L)}\})}{Z_L}, 
\end{eqnarray}
where $Z_L ( = \sum_{\{\sigma_i^{(L)}=\pm 1 \}} F_L(\{\sigma_i^{(L)}\}))$ 
is the partition function of $A_L$. 
Thus, we can determine the spin configuration $\{\sigma_i^{(L)}\}$ of the 
layer $(L)$ using a uniform random number.

The next step is to determine the spin configuration $\{\sigma_i^{(L-1)}\}$ 
of the layer $(L-1)$ under the condition that the spin configuration of 
the layer $(L)$ is given as $\{\sigma_i^{(L)}\}$. 
This is equivalent to determine the spin configuration 
$\{\sigma_i^{(L-1)}\}$ of the cluster $A_{L-1}$ which is obtained by 
removing the layer $(L)$ from $A_L$. 
The spins on this layer $(L)$ of $A_L$ now contribute to the effective 
fields on the layer $(L-1)$ of $A_{L-1}$. 
Thus, the effective field $\tilde{h}_i^{(L-1)}$ of the $i$th spin on the 
layer $(L-1)$ of $A_{L-1}$ is given as 
\begin{eqnarray}
  \tilde{h}_i^{(L-1)} = h_i^{(L-1)} + \sum_j J_{ij}^{(L-1,L)} \sigma_j^{(L)}.
\end{eqnarray}
Then, the function $F_{L-1}(\{\sigma_i^{(L-1)}\})$ of $A_L$ becomes
\begin{eqnarray}
 \tilde{F}_{L-1}(\{\sigma_i^{(L-1)}\}) &=& 
  \sum_{\{\sigma_i^{(L-2)}=\pm 1 \}} F_{L-2}(\{\sigma_i^{(L-2)}\}) 
  \exp(\beta H_{L-2,L-1} + \beta B_{L-1} + \beta H_{L-1,L}) \nonumber\\  
  &=& F_{L-1}(\{\sigma_i^{(L-1)}\})\exp( \beta H_{L-1,L}) 
\end{eqnarray}
for $A_{L-1}$. 
The probability $P_{L-1}(\{\sigma_i^{(L-1)}\})$ of the spin configuration 
$\{\sigma_i^{(L-1)}\}$ of the layer $(L-1)$ is given as
\begin{eqnarray}
 P_{L-1}(\{\sigma_i^{(L-1)}\}) 
  = \frac{\tilde{F}_{L-1}(\{\sigma_i^{(L-1)}\})}{\tilde{Z}_{L-1}} 
\end{eqnarray}
with 
\begin{eqnarray}
  \tilde{Z}_{L-1} &=& \sum_{\{\sigma_i^{(L-1)}=\pm 1 \}} 
                      \tilde{F}_{L-1}(\{\sigma_i^{(L-1)}\}) \nonumber \\
                  &=& F_L(\{\sigma_i^{(L)}\}) \exp( -\beta B_L) .
\end{eqnarray}
Thus, we can also determine the spin configuration $\{\sigma_i^{(L-1)}\}$ of 
the layer $(L-1)$. 
Repeating this procedure from the layer $(L-1)$ to the layer (1), the spin 
configuration of the cluster $A_L$ can be updated.

We can readily show that the spin configuration of $A_L$ realized in this 
procedure is in accordance with the Boltzmann weight. 
The probability of this spin configuration 
$P_{A_L}(\{\sigma_i^{(1)}\},\{\sigma_i^{(2)}\}, \cdots, \{\sigma_i^{(L)}\})$ 
is given as the product of the individual probabilities 
$P_l(\{\sigma_i^{(l)}\})$, i.e., 
\begin{eqnarray}
 P_{A_L}(\{\sigma_i^{(1)}\},\{\sigma_i^{(2)}\}, \cdots ,\{\sigma_i^{(L)}\}) 
 &=& \prod_{l=1}^L P_l(\{\sigma_i^{(l)}\}) \nonumber \\
 &=& \prod_{l=1}^{(L-1)} \frac{\tilde{F}_l(\{\sigma_i^{(l)}\})}{\tilde{Z}_{l}} 
     \times \frac{F_{L}(\{\sigma_i^{(L)}\})}{Z_L} \nonumber \\
 &=& \frac{1}{Z_L} \exp(-\beta {\cal H}_{A_L}). 
\end{eqnarray}
This is nothing but the Boltzmann weight. 
Hence we call this algorithm a cluster heat bath (CHB) method.  
When all the spins are updated one time, we call it one MC sweep just 
like in the conventional MC method.


The CHB method can be applied to clusters with any shape. It is not 
necessary that the interactions of the model are of the nearest neighbor.  
Only restriction is that we can decompose the cluster into layers as 
described in the form of eq. (2). 
It is most effective to choose the cluster as ladders with its width of $M$ 
in two-dimensions and columns with its intersection of $M_x \times M_y$ in 
three-dimensions, because we can use the transfer matrix 
technique\cite{Morgenstern}. 
The CHB method may improve relaxation in various systems. 
The updated spin configuration of the cluster depends on its environment 
but not on its original spin configuration. 
Thus, the spin structure may always fluctuate in the scale of the cluster size 
and, of course, a cluster-flip effect is automatically taken into account 
\cite{Comment1}. 
Moreover, if we choose the cluster appropriate to the model, 
we may perform most effective MC simulation. 
Here, we demonstrate it in a simulation of the ANNNI model\cite{ANNNI} in 
two-dimensions.

The ANNNI model is an array of Ising chains with ferromagnetic interaction 
$J_1$ between spins in adjacent chains and competing antiferromagnetic 
interaction $J_2$ between spins in next-nearest-neighbor chains, 
augmented by ferromagnetic nearest-neighbor interaction $J_0$ in the chains. 
It is well known that, for $\kappa \equiv -J_2/J_1 > \frac{1}{2}$, the ground 
state of the model is the $(2,\overline{2})$ antiphase described by an 
alternate arrangement of two up-spin chains and two down-spin chains, i.e., 
$\cdots + + - - + + \cdots$. 
It is believed that a floating incommensurate(IC) phase appears between 
the $(2,\overline{2})$ antiphase and the paramagnetic phase. 
However, it turned out to be much CPU time comsuming to establish the phase 
boundaries reliably\cite{ANNNI}.
The difficulty in the MC simulation of the ANNNI model is that the spin 
structure depends on initial spin configurations. 
This is because the spin structure of the IC phase near transition to the 
$(2,\overline{2})$ antiphase consists of regions of the $(2,\overline{2})$ 
antiphase separated by $+ + +$ or $- - -$ walls\cite{Selke1}. 
Therefore, it is necessary to insert 4 walls simultaneously to the 
$(2,\overline{2})$ antiphase to get the IC phase starting from the 
$(2,\overline{2})$ antiphase. 
That is, we must rearrange at least $16N_x$ spins to get the IC phase, 
where $N_x$ is the number of the spins of the chain, which is not easy 
to be realized by using the conventional single-spin-flip MC method
\cite{Comment2}. 
This difficulty is not largely relieved even when we choose an open boundary 
condition, because open boundaries lead to a pinning effect\cite{Selke2}, 
i.e., the end two chains tend to take either $+ +$ or $- -$ configuration. 
In this case, we must rearrange at least $8N_x$ spins at the ends. 
So the ANNNI model is one of the most difficult models in the computer 
simulation\cite{ANNNI}. 
However, if we use the CHB method with clusters of $M\times N_x$ spins with 
$M \geq 16$(or $M \geq 8$ at the ends), we may easily add or remove the walls.

To examine our speculation, we performed the CHB simulation of the 
model with $J_0 = J_1$ and $\kappa = 0.6$ on the lattice of $N_x\times N_y 
= 64 \times 128$ spins with open boundary conditions.  
We treated clusters of $8\times64$ spins\cite{Comment3}. At each temperature, 
we started with two different initial spin configurations, i.e., 
a paramagnetic spin configuration and the $(2,\overline{2})$ antiphase 
spin configuration, and calculated quantities of interest. 
Here we present results of the square of the chain magnetization $M_2$ which 
plays the role of the order parameter of this model\cite{Selke3,Villain}: 
\begin{eqnarray}
  M_2 = \frac{1}{N_y}\sum_j^{N_y}(\frac{1}{N_x}\sum_i^{N_x}\sigma_{ij})^2. 
\end{eqnarray}
As temperature is decreased below $T = 1.0J_1$, the relaxation becomes 
very slow.  We present, in Figs. 1(a) and 1(b), typical results of the MC 
sweep dependence of $M_2$ at $T = 0.9J_1$ in the conventional MC and CHB 
methods, respectively. 
Figure 1(a) shows most clearly the difficulty of the computer simulation 
of the ANNNI model as mentioned above. 
However, we could get its equilibrium value within a reasonable number of MC 
sweeps using the CHB method as seen in Fig. 1(b). 
We calculated the average values $\langle M_2\rangle$ at different 
temperatures using both the methods. 
Results are presented in Figs. 2(a) and 2(b) for the conventional MC method 
and the CHB method, respectively. 
At all temperatures, in the CHB method, we could get the same values starting 
with the two initial spin configurations in contrast with the conventional 
MC method\cite{Comment4}. 
Thus, we conclude that the CHB method, in fact, relieve the difficulty of the 
computer simulation of the ANNNI model.


We have proposed a new update algorithm of the spin configuration and 
demonstrated its efficiency in the ANNNI model in two-dimensions.
We should note again that the algorithm is a natural generalization 
of the conventional heat bath algorithm and applicable to various systems
with short-range interactions. Since a large fluctuation of the spin 
configuration may occur for every MC sweep, the CHB method is particularly 
useful for studying equilibrium properties of complex systems such as 
spin-glasses\cite{Shirakura}. We also note that we may perform much more 
effective MC simulation by combining the CHB method with extended ensemble 
methods such as the exchange MC method\cite{Hukushima}.

\bigskip

The authors wish to thank Dr. T. Nakamura for valuable discussions. 
They also wish to give their thanks to Dr. S. Fujiki for showing them his 
unpublished data of the ANNNI model. 
This work was partly financed by a Grant-in-Aid for Scientific Research 
from the Ministry of Education, Science and Culture. The simulations were 
made partly on FACOM VPP500 at the Institute for Solid State Physics in 
University of Tokyo.


\begin{figure}
\caption{The MC sweep dependences of the order parameter $M_2$ at 
$T = 0.9J_1$ starting with  two initial spin configurations of 
a paramagnetic phase (PARA) and the $(2,\overline{2})$ antiphase (AP) by 
(a) the conventional MC method and (b) the CHB method.}
\end{figure}

\begin{figure}
\caption{The temperature dependences of the average value of the order 
parameter $\langle M_2 \rangle$ starting with two initial spin 
configurations of a paramagnetic phase (PARA) and the $(2,\overline{2})$ 
antiphase (AP) by (a) the conventional MC method and (b) the CHB method.
In the conventional MC method, the averages were taken over $2\times10^4 
- 4\times10^4$ and $1\times10^5 - 1.5\times10^5$ MC sweeps for 
$T \geq 1.0J_1$ and  $T \leq 0.9J_1$, respectively. On the other hand, 
in the CHB method, those were taken over much smaller numbers of MC sweeps, 
i.e., $5\times10^2 - 10\times10^2$ and  $1\times10^4  - 1.5\times10^4$ 
MC sweeps for $T \geq 1.0J_1$ and  $T \leq 0.9J_1$, respectively. }
\end{figure}

\begin{references}

\bibitem{Binder} K. Binder, {\it The Monte Carlo Method in Condensed Matter 
Physics}, Second, Corrected and Updated Edition (Springier-Verlag  Berlin  
Heidelberg 1995).

\bibitem{Swendsen2} R. H. Swendsen and J. S. Wang, Phys. Rev. Lett. 
{\bf 58}, 86 (1987).

\bibitem{Wolff} U. Wolff, Phys. Rev. Lett. {\bf 62}, 361 (1989).

\bibitem{Fortuin} P. W. Kasteleyn and C. Fortuin, J. Phys. Soc. Jpn. 
Suppl. {\bf 26s}, 11 (1969). 

\bibitem{Swendsen1} R. H. Swendsen and J. S. Wang, Phys. Rev. Lett. 
{\bf 57}, 2607 (1986).

\bibitem{Berg} B. A. Berg and T. Neuhaus, Phys. Lett. B {\bf 267}, 249 (1991).

\bibitem{Marinari} E. Marinari and G. Parisi, Europhys. Lett. {\bf 19}, 
451 (1992).

\bibitem{Hukushima} K. Hukushima and K. Nemoto, J. Phys. Soc. Jpn. {\bf 65}, 
1604 (1996).

\bibitem{Koseki} O. Koseki and F. Matsubara, to appear in J. Phys. Soc. Jpn. 

\bibitem{Koseki2} O. Koseki and F. Matsubara, in preparation. 

\bibitem{Yelon} W. B. Yelon, D. E. Cox and M. Eibsch\"{u}tz, Phys. Rev. 
{\bf B12}, 5007 (1975).

\bibitem{Mekata} M. Mekata and K. Adachi, J. Phys. Soc. Jpn. {\bf 44}, 
806 (1978).

\bibitem{ANNNI} See, e.g., W. Selke, in {\it PHASE TRANSITIONS}, 
ed. C. Domb and J. L. Lebowitz (Academic Press, 1992), Vol. 15, p. 1, 
and references therein. 

\bibitem{Morgenstern} I. Morgenstern and K. Binder, Phys. Rev. Lett. 
{\bf 43}, 1615 (1979).

\bibitem{Comment1}
The CHB method is different from the cluster-flip 
methods\cite{Swendsen2,Wolff}. The term {\it cluster} in the CHB method 
denotes an ensemble of spins in a fixed part of the lattice in contrast with 
that in the cluster-flip methods in which the term is used as an ensemble 
of connected spins with the same sign. 
In the cluster-flip methods, clusters themselves are generated in accordance 
with the Boltzmann weight and the clusters may extend over the lattice. 
In the CHB method, clusters with a predetermined shape are treated and then 
the numbers of connected spins which may flip simultaneously are limited. 
However, if in each of the clusters there exist many spin configurations 
which have almost the same energy, one of them is selected in accordance 
with its Boltzmann weight. 


\bibitem{Selke1} W. Selke, K. Binder, and W. Kinzel, Surf. Sci. {\bf 125}, 74 
(1983)

\bibitem{Comment2} Since the two walls are never linked, the minimum change 
of the chain arrangement is that from $\cdots 2 \mid \overline{2}\: 2\: 
\overline{2}\: 2\: \overline{2}\: 2\: \overline{2} \: 2 \mid \overline{2} 
\cdots$ to $\cdots 2 \mid 1 \: \overline{2}\: 3 \: \overline{2}\: 2\: 
\overline{3}\: 2 \overline{1} \mid \overline{2} \cdots$.

\bibitem{Selke2} W. Selke and M. E. Fisher, Phys. Rev. B {\bf 20} 257 (1979). 

\bibitem{Comment3} By using the CHB method with $M = 8$, we can readily 
add the walls at the ends and put them inside the lattice and vice versa. 

\bibitem{Selke3} W. Selke, Z. Physik B {\bf 43}, 335 (1981). 

\bibitem{Villain} J. Villain and P. Bak, J. Physique {\bf 42}, 657 (1981). 

\bibitem{Comment4} In a standard UNIX machine, we need about 30 times much 
CPU time per one MC sweep in the CHB method with $M = 8$ than that in the 
conventional single-spin-flip method. However, we only need about 
one-hundredths of MC sweeps even for $T > 1.0J_1$, to get similar accuracy of 
data, and the ratio becomes much smaller at lower temperatures. 

\bibitem{Shirakura} T. Shirakura and F. Matsubara, in preparation.

\end{references}
\end{document}